\documentclass[final,5p,times,twocolumn]{elsarticle}

\usepackage{amssymb}
\usepackage{amsmath}
\usepackage{bm}

\journal{Science Bulletin}

\begin{document}

\begin{frontmatter}



\title{Gyrotropic Magnetic Effect in Black Phosphorus Irradiated with Bicircular Light}

\author[1,2]{Fangyang Zhan}
\author[3]{Xin Jin}
\author[1,4]{Da-Shuai Ma}
\author[5]{Jing Fan}
\author[3]{Peng Yu}
\author[1,4]{Dong-Hui Xu\corref{cor}}\ead{donghuixu@cqu.edu.cn}
\author[1,4]{Rui Wang\corref{cor}}\ead{rcwang@cqu.edu.cn}
\cortext[cor]{Corresponding authors.}
\affiliation[1]{organization={Institute for Structure and Function $\&$ Department of Physics $\&$ Chongqing Key Laboratory for Strongly Coupled Physics, Chongqing University}, 
            city={Chongqing 400044},
%            postcode={400044},
            country={China}}
\affiliation[2]{organization={College of Mathematics and Statistics, Chongqing University}, 
            city={Chongqing 400044},
%            postcode={400044},
            country={China}}            
\affiliation[3]{organization={College of Physics and Electronic Engineering, Chongqing Normal University}, 
            city={Chongqing 400044},
%            postcode={400044},
            country={China}}
\affiliation[4]{organization={Center of Quantum materials and devices, Chongqing University}, 
            city={Chongqing 400044},
%            postcode={400044},
            country={China}}
\affiliation[5]{organization={Center for Computational Science and Engineering, Southern University of Science and Technology},
            city={Shenzhen 518055},
%            postcode={518055}, 
            country={China}}

%\begin{abstract}
%\end{abstract}

%\includegraphics{Graphical abstract.pdf}

\end{frontmatter}



Recent years have witnessed a surge of interest in topological semimetals due to their unique electronic band structures and exotic quantum phenomena~\cite{RevModPhys.90.015001}. 
Among them, Weyl semimetals (WSMs) host massless chiral fermions as low-energy excitations~\cite{PhysRevB.83.205101}, leading to novel transport phenomena, such as the chiral magnetic effect (CME)~\cite{PhysRevD.78.074033}, which arises from the chiral anomaly and results in a nonequilibrium current parallel to an applied magnetic field when an electric field is also present. However, a static magnetic field alone cannot induce a current; that is, the CME in equilibrium is zero.

A less explored but fundamentally distinct response is the gyrotropic magnetic effect (GME), which can occur in inversion ($\mathcal{P}$) symmetry broken WSMs under a dynamic magnetic field. Unlike the CME, the GME originates from the orbital magnetic moment of Bloch electrons and is directly proportional to the energy separation between Weyl nodes with opposite chirality~\cite{PhysRevLett.116.077201}. While the GME represents a basic property of Bloch electrons in a system with $\mathcal{P}$ symmetry breaking, it remains unexplored experimentally. The obstacle to measurements of GME possibly attributes to the fact that $\mathcal{P}$-symmetry broken WSMs typically possess time-reversal ($\mathcal{T}$) symmetry. The combination of $\mathcal{T}$ symmetry and additional crystalline symmetries enforces that the energy difference between Weyl nodes with opposite chirality is zero, naturally vanishing the gyrotropic current in these $\mathcal{T}$-invariant WSMs. The intrinsic chiral semimetals, such as CoSi and AlPt, which host topological nodes at different energies, are natural candidates for the GME. However, these materials typically possess multiple pairs of Weyl nodes near the Fermi level, and the resulting superposition of their individual GME contributions can complicate the interpretation and direct detection of the effect. To avoid such mishaps and ensure the observation of a visible gyrotropic current, it is highly desirable to design an ideal WSM candidate that hosts a single pair of Weyl nodes with a sizable energy difference near the Fermi level.

Floquet engineering has emerged as a promising approach for designing exotic topological states through the utilization of time-periodic light fields~\cite{Light2022,2020nonequilibrium}. In particular, employing the irradiation of bicircular light (BCL) with an integer frequency ratio has been attracting increasing interest due to its enhancement of Floquet engineering capabilities~\cite{2024valleytronics,2024BN}. The BCL irradiation can break not only $\mathcal{T}$ symmetry but also selectively break spatial symmetries, which enables advanced manipulation on material properties~\cite{PhysRevLett.127.126601,lesko2025}. Even so, the possibility of generating an energy difference between light-driven Weyl nodes in nodal-line semimetals (NLSMs) has remained unrecognized, leaving the associated novel physical phenomena unexplored. 

Here, we theoretically propose a robust scheme to generate a significant GME in anisotropic nodal-line semimetals using Floquet engineering with BCL. We show that BCL irradiation can selectively break spatial and $\mathcal{T}$ symmetries, inducing a topological phase transition from a NLSM to a WSM with a minimal number of Weyl nodes. Compared with the intrinsic $\mathcal{P}$-symmetry broken WSMs, our proposal offers a key practical advantage, i.e., controlling the degree of symmetry breaking, thereby enabling the magnitude of the gyrotropic current to oscillate with BCL polarization and facilitating the identification of this unique GME current signal. Using first-principles calculations combined with Floquet theory, we identify compressed black phosphorus (BP) as an ideal material platform. The inherent crystal anisotropy and the absence of particle-hole symmetry in realistic NLSMs can lead to paired Weyl nodes with a sizable energy separation.
Given the significant achievements in Floquet band engineering of BP~\cite{2023Pseudospin,Manipulating2024,PhysRevLett.120.237403,PhysRevLett.131.116401}, we expect that our results can be experimentally detected using laser-triggered photoconductive switches and will attract considerable attention.

\begin{figure*}
    \centering
    \includegraphics[width=1.0\linewidth]{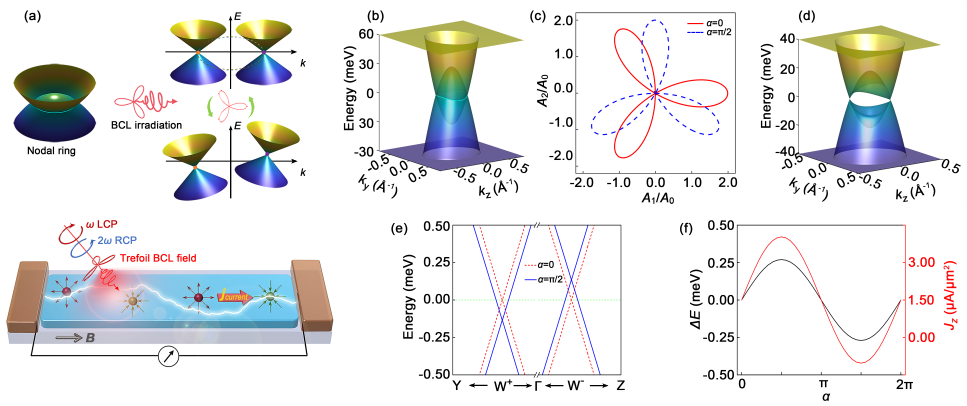}
    \caption{The conceptual illustration and model results of light-dressed an anisotropic nodal line semimetal. 
    (a) The conceptual illustration and proposed setup for a gyrotropic current driven by trefoil BCL.
    (b) A nodal ring with varying energy is located at the $k_y-k_z$ plane before application of light to the system.  
    (c) Two-frequency Lissajous curves for $\alpha=0$ and $\pi/2$. 
    (d) The transition from a NLSM to WSM under illumination with BCL. 
    (e) The band structures around two Weyl nodes for different BCL polarization state $\alpha$. 
    (f) The BCL-controlled energy separation (black line) between Weyl nodes and the magnitude of the gyrotropic current (red line). 
    The values of parameters are $C=0.03$ eV, $D=0.05$ eV \AA$^{2}$, $v=0.4$ eV \AA$^{-1}$, $\epsilon_1=- \epsilon_2=0.45$ eV \AA$^{-4}$, $eA_0/\hbar = 0.1$ \AA$^{-1}$, $\hbar\omega=1$ eV, $\alpha=\pi/2$, and $\varphi=\pi/4$. We fix magnetic field in the $z$ direction with amplitude of 3 T.
    \label{FIG1}}
\end{figure*}

Considering the BCL irradiation generated by interferometrically combining circularly polarized light at frequency $\omega$ with its counter-rotating higher-harmonic $\eta\omega$. It can be expressed as
\begin{equation}\label{Eq.1}
\mathbf{A}(t)={A}_0\sqrt{2}\mathrm{Re}[e^{-i(\eta\omega t-\alpha)}\boldsymbol{\varepsilon}_R+e^{-i\omega t}\boldsymbol{\varepsilon}_L],
\end{equation}
where $A_0$ is the amplitude of right-handed circularly polarized (RCP) and left-handed circularly polarized (LCP) light, $\eta$ and $\alpha$ are respectively the frequency ratio and phase difference between RCP and LCP light. $\varepsilon_{R(L)}$ is RCP (LCP) light polarization in the plane of two orthonormal unit vectors, as shown in the supplementary material (SM).
As shown in Fig.~\ref{FIG1}a, when the relative phase $\alpha$ of BCL is set to 0 or $\pi$, an additional joint symmetry $\mathcal{MT}$ is imposed on the paired Weyl nodes with opposite chirality, leading to these symmetry-protected Weyl nodes being pinned at the same energy. For all other values of $\alpha$, they reside in different energy due to the breaking  $\mathcal{MT}$ symmetry, thereby a gyrotropic current is present when a slowly oscillating magnetic field is applied. 

To reveal the origin of the gyrotropic current driven by BCL, we start from a nodal line system, and the physics in the absence of light irradiation can be described by a two-band Hamiltonian~\cite{PhysRevLett.117.087402}.
\begin{equation}\label{Eq.2}
H\left(\mathbf{k}\right)=P_k\sigma_z+Q_k\sigma_x+I_k\sigma_0,
\end{equation}
where $P_k=C-Dk^2$, $Q_k=vk_x+ \lambda k^{3}_x$,  $I_k=\epsilon_1k^{4}_y+\epsilon_2k^{4}_z$, and $k^{2}=k^{2}_x+k^{2}_y+k^{2}_z$. $C$ and $D$ are model parameters. $v$ is the Fermi velocity along the $x$ direction, and $\epsilon_1$ and $\epsilon_2$ represent anisotropic energy parameters in the $k_y$-$k_z$ plane. We can easily find that the presence of a nodal ring on the $k_x=0$ plane is determined by the equation $k^{2}_y+k^{2}_z=C/D$ and is protected by mirror symmetry $\mathcal{M}_x$. The $I_k$ arises from crystal anisotropy and the absence of particle-hole symmetry, which has no influence on the nodal ring equation but causes a variation in the energy along the nodal line (Fig.~\ref{FIG1}b).

\begin{figure*}
    \centering
    \includegraphics[width=1.0\linewidth]{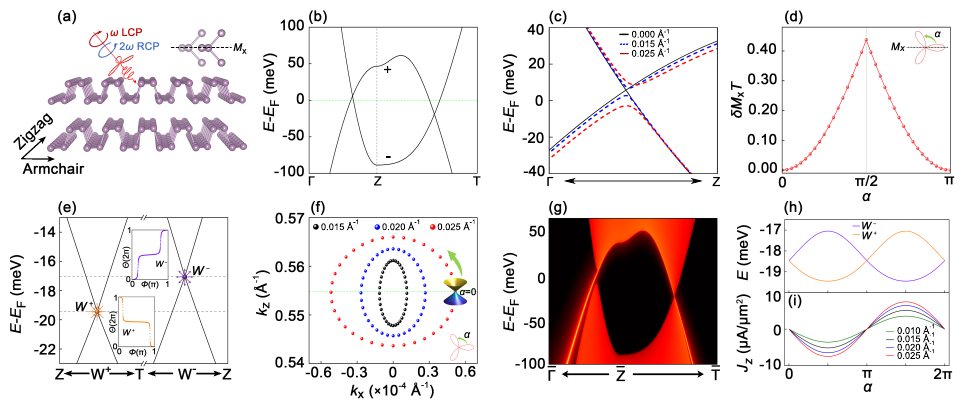}
    \caption{The electronic structure evolution and gyrotropic current induced by trefoil BCL in compressed BP. 
    (a) The atomic structure of BP and a schematic for irradiation of BCL field. 
    (b) Without light irradiation, the nodal ring in the compressed BP exhibits a variation in energy relative to the Fermi level. 
    (c) Evolution of electronic band structures of compressed BP under the irradiation of BCL, and the band gap is enlarged with an increase of light intensity. 
    (d) The degree of $\mathcal{M}_x\mathcal{T}$ breaking under the BCL irradiation. 
    (e) The enlarged view of band structure around two Weyl nodes is shown to emphasize the energy separation. The insets show the evolution of the Wannier charge centers around the $W^{+}$ and $W^{-}$, respectively. 
    (f) Trajectory of the Weyl nodes (marked by the filled dots) in momentum space as $\alpha$ evolves from 0 to $2\pi$ by rotating the light waveform.
    (g) The calculated surface state projected on the semi-infinite (100) surface of compressed BP under the irradiation of BCL.
    (h) The energy of Weyl nodes as a function of BCL with polarization state $\alpha$.
    (i) The gyrotropic current at different light intensities.
    For (e), (g), (h) and (i), we set the light intensity $eA_0/\hbar = 0.025$ \AA$^{-1}$, incident angle $\varphi=\pi/4$, and magnetic field in the $z$ direction with amplitude of 1 T.
    \label{FIG2}}
\end{figure*}

For concreteness, we focus on the effect of trefoil BCL, where $\eta=2$. The detailed BCL field settings are shown in SM. To align the symmetries of BCL with the system, we provide the Lissajous curves of the BCL for $\alpha=0$ and $\pi/2$, as shown in Fig.~\ref{FIG1}c. It can be observed that the $\mathcal{M}_x\mathcal{T}$ symmetry is preserved only when one of the arms of the trefoil pattern aligns with the $x$ axis. The coupling of the light field to the system is introduced via the minimal coupling substitution in the Hamiltonian $H(\mathbf{k},t)= H[\mathbf{k}+\mathbf{A}(t)]$. In our work, the physics is dominated by band inversion near the Fermi level around the nodal line. Since we only need to focus on the low-energy regime near the nodes, the off-resonant laser frequency $\hbar \omega = 1$ eV (much larger than the band inversion scale) can satisfy the high-frequency approximation. In this case, one can obtain a static effective Floquet-Bloch Hamiltonian~\cite{PhysRevLett.117.087402,PhysRevLett.128.066602}:
\begin{equation}\label{Eq.3}
\begin{split}
H_{\mathrm{eff}}(\mathbf{k})&=H^0(\mathbf{k})+\sum_{n\geq1}\frac{[H^{n},H^{-n}]}{n\hbar\omega}+\mathcal{O}(\omega^{-2}), \\
&=\tilde{P}_k\sigma_z+\tilde{Q}_k\sigma_x+\tilde{M}_k\sigma_y+\tilde{I}_k\sigma_0.
\end{split}
\end{equation}
Here, $H^n(\mathbf{k})=(1/T)\int_0^T\mathrm{~d}te^{-in\omega t}H[\mathbf{k}+\mathbf{A}(t)]$ is the $n$th Fourier component of the time-dependent Hamiltonian, where $T$ is the period of light. The light-renormalized coefficients of the Pauli matrix and the identity matrix can be expressed in SM. The primary effect of driving the NLSM state (Fig.~\ref{FIG1}b) to the WSM state (Fig.~\ref{FIG1}d) manifests in the emergence of a momentum-dependent $\tilde{M}_k\sigma_y$ term, which is absent in the original Hamiltonian (i.e., Eq.~\ref{Eq.2}). Crucially, when $\alpha$ is set to 0, we can observe a pair of Weyl nodes with energy degeneracy due to protection of the $\mathcal{M}_x\mathcal{T}$ symmetry. In contrast, when $\alpha = \pi/2$, this symmetry is broken, and the energy degeneracy is lifted (Fig.~\ref{FIG1}e). The GME response coefficient $\gamma_{\nu}^{\rm GME}$ is derived in detail in SM. The gyrotropic current under a slowly varying magnetic field $\bm{B}$ is given by
\begin{equation}\label{Eq.4}
\bm{J}=\frac{2e^2}{h^2} A_0^3 R\sin\alpha (\epsilon_1 \sin\varphi\cos^3\varphi -\epsilon_2 \cos\varphi\sin^3\varphi)\bm{B},
\end{equation}
where $R=\sqrt{C/D-2A_0^2-q_{x}^2}$ with the $x$ coordinate $q_x$ of the partner Weyl nodes. The Eq.~\ref{Eq.4} encapsulates the entire physics, i.e., the current is driven by the magnetic field $\bm{B}$, enabled by the BCL-induced Weyl node splitting, and amplified by the intrinsic material anisotropy (Fig.~\ref{FIG1}f).
%Unlike other potential photocurrents, the BCL-induced gyrotropic current provides unambiguous signatures: it shows a linear dependence on the magnetic field, a distinctive cubic scaling with the light field amplitude, and oscillates with BCL polarization.

Next, we demonstrate the emergence of the gyrotropic current in a realistic material, namely compressed BP. As shown in Fig.~\ref{FIG2}a, bulk BP exhibits a van der Waals layered structure with alternating stacking order. A compressive strain can induce a topological transition from a trivial semiconductor to a nodal-line semimetal, which has been experimentally confirmed~\cite{PhysRevLett.115.186403,PhysRevB.109.L201103}.
Without loss of generality, we select a moderate compressive strain of 3.5\% along the armchair direction in our work. The calculated details are included in SM. Without light irradiation, two bands with opposite parity are inverted around the $\rm Z$ point near the Fermi level (Fig.~\ref{FIG2}b). As a result, the nontrivial band topology can lead to the existence of a nodal ring in the $\Gamma$-$\rm Z$-$\rm T$ plane of the Brillouin zone due to $\mathcal{M}_x$ symmetry. Under the irradiation of BCL field, where the polarized component of light is along the armchair direction, a light-driven topological transition occurs from the NLSM to a WSM phase, forming a single pair of Weyl nodes. Due to the crystal anisotropy and the absence of particle-hole symmetry in realistic material systems, the nodal ring exhibits a non-flat dispersion, as shown in Fig.~\ref{FIG2}b. Once light intensity $eA_0/\hbar \neq $0 {\AA}$^{-1}$, the band structure immediately becomes gapped except at the two Weyl nodes, and the band gap is enlarged as the light intensity increases (Fig.~\ref{FIG2}c).

To further demonstrate the unique capability of BCL, we calculate the degree of $\mathcal{M}_x\mathcal{T}$ breaking under trefoil BCL irradiation. The breaking of $\mathcal{M}_x\mathcal{T}$ symmetry can be generally assessed by evaluating the overlap between the field $\mathbf{A}(t)$ and its symmetry-transformed counterpart~\cite{acsphotonics}. As shown in Fig.~\ref{FIG2}d, the symmetry breaking is maximized when the polarization state $\alpha = \pi/2$, at which point a pair of light-induced maximally misaligned Weyl nodes emerges (Fig.~\ref{FIG2}e). The topological nature is confirmed via the Wilson loop method, applied to track the evolution of the Wannier charge center (insets in Fig.~\ref{FIG2}e).
To understand the evolution of Weyl nodes under BCL driving, we trace their positions as functions of $\alpha$ and light amplitude. In Fig.~\ref{FIG2}f, the nodes form closed trajectories in momentum space as $\alpha$ varies from 0 to $2\pi$, with the loop expanding as light intensity increases.
To reveal the resulting nontrivial band topology, we calculate the photon-dressed surface local density of states on the (100) surface. As shown in Fig.~\ref{FIG2}g, a clear gap is visible along $\rm\bar{Z}$-$\bar{\Gamma}$ and a projected Weyl cone with linear dispersion along $\rm\bar{Z}$-$\rm\bar{T}$.

Furthermore, as shown in Fig.~\ref{FIG2}h, the energy shift of the Weyl nodes depends sensitively on the polarization parameter $\alpha$. Specifically, the two BCL-induced Weyl nodes connected by $\mathcal{M}_x\mathcal{T}$ symmetry remain energetically degenerate at $\alpha = 0$ or $\pi$. For other values of $\alpha$, however, $\mathcal{M}_x\mathcal{T}$ symmetry is broken, lifting the degeneracy between the oppositely chiral Weyl nodes, which is consistent with the previous low-energy effective model. This controllability enables the manipulation of a gyrotropic current by applying a slowly oscillating magnetic field. As demonstrated in Fig.~\ref{FIG2}i, both the polarization and amplitude of the trefoil BCL can continuously modulate this current, offering wide-range and switchable control in compressed BP. Notably, under a light intensity of $eA_0/\hbar = 0.01$~\AA$^{-1}$, the current reaches experimentally accessible magnitudes on the order of $\mu$A/$\mu$m$^2$. Different from BCL-irradiated Dirac semimetals~\cite{PhysRevLett.128.066602}, the energy separation of paired Weyl nodes in the NLSM is dominated by its intrinsic non-flat dispersion, resulting from crystal anisotropy and the lack of particle-hole symmetry. This crucial feature leads to a significantly enhanced gyrotropic current in compressed BP, three orders of magnitude larger than that predicted in the Dirac semimetal Cd$_3$As$_2$~\cite{PhysRevLett.128.066602}.

Finally, to facilitate experimental observation, we assess the feasibility of detecting the gyrotropic current. In terms of the parameters for inducing the gyrotropic current, the pump photon energy and amplitude of BCL are set as $\hbar \omega = 1$ eV and $eA_0/\hbar = 0.01$ \AA$^{-1}$, corresponding to an electric field peak strength of $9.96 \times 10^7$ V$/$m (or energy density of $2.63 \times 10^{9}$ W$/$cm$^2$). In the previous time- and angle-resolved photoelectron spectroscopy experimental study~\cite{2023Pseudospin}, the penetration depth was estimated to be $\sim14$ nm, corresponding to a stack of 25-30 atomic layers. As the used probe depth is significantly smaller than $\sim3.5$ nm, this implies that the region being measured experiences a nearly uniform pump fluence. Therefore, it is experimentally feasible to create a sufficiently thick, uniformly illuminated region where the predicted topological phase transition and the associated bulk GME can be realized and measured. Besides, a pulse duration of 90 fs is sufficient to realize Floquet band renormalization in BP~\cite{2023Pseudospin} at a low fluence of 0.24 mJ/cm$^2$, well below the laser damage threshold~\cite{adma.201704405}, thereby minimizing heating. Moreover, even at higher intensities where heating occurs, the gyrotropic current can be isolated via lock-in measurements, as spurious signals and heating are independent of $\alpha$. Experimentally, the GME current vanishes at zero magnetic field and oscillates with BCL polarization, allowing it to be clearly distinguished from other transport phenomena. These advantages position compressed BP under BCL irradiation as a highly promising platform for observing and controlling the gyrotropic current.

In summary, based on the low-energy effective model, we propose a robust scheme to generate significant GME in anisotropic NLSM using Floquet engineering with BCL. The BCL irradiation can selectively break spatial and $\mathcal{T}$ symmetries, inducing a topological phase transition from a NLSM to a WSM with a misaligned minimal number of Weyl nodes. Therefore, the photoinduced gyrotropic current is naturally present according to the GME if an additional slowly varying magnetic field is applied.
Using first-principles calculations combined with Floquet theory, we identify compressed BP as an ideal material platform. The intrinsic anisotropy of BP amplifies the GME, resulting in a measurable gyrotropic current that is several orders of magnitude larger than that in previously predicted Cd$_3$As$_2$~\cite{PhysRevLett.128.066602}.
Our work not only provides a concrete path toward the experimental realization of GME but also opens new avenues for exploring the interplay of light, symmetry, and topology in quantum materials.

\section*{Conflict of interest}
The authors declare that they have no conflict of interest.

\section*{Acknowledgements}
This work was supported by the National Natural Science Foundation of China (12204074, 12222402, 92365101, 12347101, 12074108, 12447141, 12404045, and 12474151), the Natural Science Foundation of Chongqing (2023NSCQ-JQX0024 and CSTB2022NSCQ-MSX0568), the Beijing National Laboratory for Condensed Matter Physics (2024BNLCMPKF025), the Postdoctoral Fellowship Program of CPSF (GZC20252254), the Special Funding for Postdoctoral Research Projects in Chongqing (2024CQBSHTB2036), and the Science and Technology Research Program of Chongqing Municipal Education Commission (KJZD-K202500512, KJQN-202400553).

\section*{Author contributions}
Rui Wang and Dong-Hui Xu supervised the work. Fangyang Zhan and Xin Jin performed the theoretical calculations. Rui Wang, Dong-Hui Xu, Da-shuai Ma, jing Fan, and Peng Yu supported the data analysis and helped interpret the results. Fangyang Zhan, Dong-Hui Xu, and Rui Wang prepared the manuscript with comments from all authors. All authors have approved the manuscript.

\appendix
\section{Supplementary material}
\label{app1}

\bibliographystyle{elsarticle-num} 
%\bibliography{ref.bib}

\begin{thebibliography}{10}
\expandafter\ifx\csname url\endcsname\relax
  \def\url#1{\texttt{#1}}\fi
\expandafter\ifx\csname urlprefix\endcsname\relax\def\urlprefix{URL }\fi
\expandafter\ifx\csname href\endcsname\relax
  \def\href#1#2{#2} \def\path#1{#1}\fi

\bibitem{RevModPhys.90.015001}
Armitage NP, Mele EJ, Vishwanath A. Weyl and Dirac semimetals in three-dimensional solids. Rev Mod Phys 90 (2018) 015001.

\bibitem{PhysRevB.83.205101}
Wan XG, Turner AM, Vishwanath A, et al. Topological semimetal and Fermi-arc surface states in the electronic structure of pyrochlore iridates. Phys Rev B 83 (2011) 205101.

\bibitem{PhysRevD.78.074033}
Fukushima K, Kharzeev DE, Warringa HJ. Chiral magnetic effect. Phys Rev D 78 (2008) 074033.

\bibitem{PhysRevLett.116.077201}
Zhong S, Moore JE, Souza I. Gyrotropic magnetic effect and the magnetic moment on the Fermi surface. Phys Rev Lett 116 (2016) 077201.

\bibitem{Light2022}
Bao CH, Tang PZ, Sun D, et al, Light-induced emergent phenomena in 2D materials and topological materials, Nat Rev Phys 4 (2022) 33.

\bibitem{2020nonequilibrium}
Rudner MS, Lindner NH. Band structure engineering and non-equilibrium dynamics in Floquet topological insulators. Nat Rev Phys 2 (2020) 229-244.

\bibitem{2024valleytronics}
Tyulnev I, Jimenez-Galan A, Poborska J, et al. Valleytronics in bulk $\mathrm{MoS}_2$ with a topologic optical field. Nature 628 (2024) 746.

\bibitem{2024BN}
Mitra S, Jimenez-Galan A, Aulich M, et al. Light-wave-controlled Haldane model in monolayer hexagonal boron nitride. Nature 628 (2024) 752.

\bibitem{PhysRevLett.127.126601}
Neufeld O, Tancogne-Dejean N, Giovannini UD, et al. Light-driven extremely nonlinear bulk photogalvanic currents. Phys Rev Lett 127 (2021) 126601.

\bibitem{lesko2025}
Lesko DMB, Weitz T, Wittigschlager S, et al. Optical control of electrons in a Floquet topological insulator.
\newblock \href {http://arxiv.org/abs/2407.17917} {\path{arXiv:2407.17917} (2024)}.

\bibitem{2023Pseudospin}
Zhou SH, Bao CH, Fan BH, et al. Pseudospin-selective Floquet band engineering in black phosphorus. Nature 614 (2023) 75.

\bibitem{Manipulating2024}
Bao CH, Schuler M, Xiao T, et al. Manipulating the symmetry of photon-dressed electronic states. Nat Commun 15 (2024) 10350.

\bibitem{PhysRevLett.120.237403}
Liu H, Sun JT, Cheng C, et al. Photoinduced nonequilibrium topological states in strained black phosphorus. Phys Rev Lett 120 (2018) 237403.

\bibitem{PhysRevLett.131.116401}
Zhou SH, Bao CH, Fan BH, et al. Floquet engineering of black phosphorus upon below-gap pumping. Phys Rev Lett 131 (2023) 116401.

\bibitem{PhysRevLett.117.087402}
Yan Z, Wang Z, Tunable Weyl points in periodically driven nodal line semimetals. Phys Rev Lett 117 (2016) 087402.

\bibitem{PhysRevLett.128.066602}
Trevisan TV, Arribi PV, Heinonen O, et al. Bicircular light Floquet engineering of magnetic symmetry and topology and its application to the Dirac semimetal {${\mathrm{Cd}}_{3}{\mathrm{As}}_{2}$}. Phys Rev Lett 128 (2022) 066602.

\bibitem{PhysRevLett.115.186403}
Xiang ZJ, Ye GJ, Shang C, et al. Pressure-induced electronic transition in black phosphorus. Phys Rev Lett 115 (2015) 186403.

\bibitem{PhysRevB.109.L201103}
Akiba K, Akahama Y, Tokunaga M, et al. Realization of nodal-ring semimetal in pressurized black phosphorus. Phys Rev B 109 (2024) L201103.

\bibitem{acsphotonics}
Neufeld O. Degree of time-reversal and dynamical symmetry breaking in electromagnetic fields and its connection to Floquet engineering. ACS Photonics 12 (2025) 2151-2159.

\bibitem{adma.201704405}
Qiu G, Nian Q, Motlag M, et al. Ultrafast laser-shock-induced confined metaphase transformation for direct writing of black phosphorus thin films. Adv Mater 30 (2018) 1704405.

\end{thebibliography}

\end{document}